\documentclass[12pt]{article}
\usepackage{epsf}
\usepackage{graphicx}
\font\af=msbm12

\newcommand{\eqnn}[1]{\begin{eqnarray*}#1\end{eqnarray*}}
\newcommand{\eqnl}[2]{\par\parbox{14.5cm}
{\begin{eqnarray*}#1\end{eqnarray*}}\hfill
\parbox{1cm}{\begin{eqnarray}\label{#2}\end{eqnarray}}}

\newcommand{\eqngrlb}[3]{\par\parbox{12.5cm}
{\begin{eqnarray}\fbox{$\displaystyle
#1\\#2$}\end{eqnarray}}\hfill
\parbox{1cm}{\begin{eqnarray}\label{#3}\end{\eqnarray}}}

\newcommand{\refs}[1]{(\ref{#1})}
\def\xin{\xi^0}
\def\vrho{\vec{\rho}}

\def\vx{\vec{x}}
\def\intl{\int\limits}
\def\di{\displaystyle}
\def\lam{\lambda}
\def\&{&\di}
\def\bg{\begin{eqnarray}\begin{array}{rcl}\displaystyle}
\def\eg{\end{array} &\di    &\di   \end{eqnarray}}
\def\bm#1{\begin{eqnarray}\begin{array}{#1}\di} 
\def\bmo#1{\begin{eqnarray*}\begin{array}{#1}\di} 
\def\eg{\end{array} &\di    &\di   \end{eqnarray}}
\def\bgo{\begin{eqnarray*}\begin{array}{rcl}\displaystyle}
\def\ego{\end{array} &\di    &\di \nonumber  \end{eqnarray*}}

\def\btensor#1#2{\renew\left#1\begin{array}{#2}\di}
\def\etensor#1{\end{array}\right#1}

\def\d{{\mbox d}}

\def\tr{\mbox{tr}\,}

\def\T{{\mbox T}}

\def\id{1\!\mbox{l}}

\def\ov{\over}
\def\qinst{q}

\def\vex0{\vx _{\rm d}}

\def\pa{\partial}
\def\al{\alpha}

\def\R{\mbox{\af R}}

\def\T{\mbox{\af T}}

\def\Z{\mbox{\af Z}}
\def\CA{{\cal A}}

\def\CD{{\cal D}}

\def\CF{{\cal F}}

\def\CH{{\cal H}}

\def\CP{{\cal P}}
\def\CS{{\cal S}}

\def\pan{\par\noindent}

\def\mus{\mu_{(\sigma)}}
\def\als{\al_{(\sigma)}}

\def\ali{{\al_{(i)}}}
\def\alj{{\al_{(j)}}}

\def\muj{{\mu_{(j)}}}

\def\rank{\,\mbox{rank}\,}
\def\dim{\,\mbox{dim}\,}
\newcommand{\mtxt}[1]{\quad\hbox{{#1}}\quad}

\date{\today}

\def\renew{\renewcommand{\arraystretch}{1}}

\begin{document}

\begin{flushright}
FSUJ-TPI-98/13\\
November 1998\\
hep-th/9811248\\     
\end{flushright}
\par
\vskip .5 truecm

\large \centerline{ 
$SU(N)$-Gauge Theories in Polyakov Gauge on the Torus\footnote{Supported 
by the Deutsche Forschungsgemeinschaft,
DFG-Wi 777/3-2}} 
\vskip 1truecm
\normalsize

\begin{center}
\textbf{C.~Ford\footnote{
Present address:  DESY,
 Platanenallee 6, D-15738 Zeuthen, Germany. e-mail: ford@ifh.de}
, T.~Tok\footnote{e-mail: Tok@tpi.uni-jena.de}
 and A.~Wipf\footnote{ e-mail: Wipf@tpi.uni-jena.de}}\\
\it{Theor.--Phys. Institut, Universit\"at Jena\\ 
Fr\"obelstieg 1, D--07743 Jena, Germany}
\end{center}
\par
\vskip 1 truecm
\begin{abstract} 
We investigate the Abelian projection with respect
to the Polyakov loop operator for
$SU(N)$ gauge theories on the four torus. The gauge fixed
$A_0$ is time-independent and diagonal. 
We construct fundamental domains for $A_0$. 
In sectors with non-vanishing instanton number such gauge fixings
are always singular. The singularities define
the positions of magnetically charged monopoles, strings
or walls. These magnetic defects sit on the Gribov horizon and
have quantized magnetic charges.
We relate their magnetic charges to the instanton number.
\end{abstract}




\vskip .5 truecm

\noindent%
In the absence of dynamical fermions the
relevant observables for confinement
studies are products of Wilson-loops \cite{wilson}.
At finite temperature $T=1/\beta$ the gauge potentials
in the functional integral
are periodic in Euclidean time i.e.
\eqnn{
A(x^0+\beta,\vx)=A(x^0,\vx)\, , \quad A (x) = A_\mu (x) \, d x^\mu}
and one may use Polyakov loops \cite{polloops}
\eqnl{
P(\vx)=\tr \big(
\CP (\beta,\vx)\big),
\mtxt{where}\CP (x^0,\vx)=\CP \exp\left[
i\int^{x^0}_0 d\tau A_0(\tau,\vx)\right]}{defpol}
as order parameters for confinement. Below
we set $\CP(\beta,\vx)\equiv \CP(\vx)$.\pan
We shall follow the strategy put forward by G. 't Hooft
\cite{tH} who considered Yang-Mills theories on a
Euclidean space-time torus $\T^4$. The torus provides a
gauge invariant infrared cut-off. Its
non-trivial topology gives rise to a non-trivial structure
in the space of Yang-Mills fields which yields additional
information on the possible phases of Yang-Mills theories.

Since the gauge invariant $P(\vx)$ 
is a functional of $A_0$ only,
we seek a gauge fixing where $A_0$ is as simple as 
possible. 
In an earlier paper \cite{us} we considered
an Abelian projection where the gauge-fixed $A_0$
is time-independent and diagonal.
The fixing hinges on the  diagonalization
of the path ordered exponential $\CP(\vx)$.
In the topologically non-trivial sectors the gauge fixed potential
has unavoidable singularities \cite{tHooft,schierholz}.
These singularities, which can be interpreted as magnetically charged
`defects',  occur at points (or loops and sheets)
where ${\cal P}(\vx)$ has degenerate eigenvalues.
For $SU(2)$ this happens when $\CP(\vx)=\pm\id$.
Associated with the gauge fixing procedure one
can define an Abelian magnetic potential $A_{mag}$ \cite{tHooft} on $\T^3$.
This allows us to precisely define the magnetic charge of
any defect.
With the gauge group $SU(2)$ the possible magnetic charges are
quantized.
While the total magnetic charge on $\T^3$ is zero, the
total magnetic charge of $\CP(\vx)=\id$ defects is equal to the
instanton number $q$  \cite{reinhardt,us,Lenz3}.
The relationship between magnetic charge and the instanton number
was considered earlier in a different context by
 Christ and Jackiw \cite{Jackiw}
and Gross et.al \cite{pisarski}.

In this letter we extend our results to $SU(N)$ and show that the
defects sit on the Gribov horizon.
Now the magnetic potential $A_{mag}$, and hence the magnetic charge
$Q_M$ are diagonal matrices.
We find that there are $N$ types of \textit{ basic defects }
corresponding to the $N$ boundary faces of the
\textit{ fundamental domain } for the gauge fixed
$A_0$.
For a basic defect, $Q_M$ is
an integer multiple of a fixed matrix.
Much as in the $SU(2)$ analysis there is a simple linear relation
between the total magnetic charge
of a given type of defect and the instanton number $\qinst$.
We again  have overall magnetic charge neutrality on $\T^3$.

We view the four torus as $\R^4$ modulo the lattice generated by
four orthogonal vectors $b_\mu,\;\mu=0,1,2,3$.
The Euclidean lengths of the $b_\mu$ are denoted by
$L_\mu$ (we may identify $L_0$ with the inverse temperature $\beta$).
Local gauge invariants such as $\tr\,F_{\mu \nu} F_{\mu \nu}$
are periodic with respect to a shift by an arbitrary lattice vector. 
However, the gauge potential $A$ has to be periodic only up to gauge 
transformations. In order to specify boundary conditions for $A$
on the torus one requires a set of $SU(N)$ valued transition
functions $U_\mu(x)$, which are defined on the whole of $\R^4$.
The periodicity properties of $A$ are as follows
\eqnn{
A(x+b_\mu)
=U_\mu^{-1}(x)A(x)U_\mu(x)+i U_\mu^{-1}(x)\d U_\mu(x),
\qquad \mu=0,1,2,3\,.}
It follows at once, that the path ordered exponential
$\CP (x^0,\vx)$ in \refs{defpol} has the
following periodicity properties 
\eqnl{
\CP(x^0\!+\!L_0,\vx )=\CP(x^0,\vx )\CP (L_0,\vx)\,,\quad\quad
\CP (x^0,\vx \!+b_i)=U_i^{-1}(x^0,\vx )\CP (x^0,\vx )
U_i(0,\vx)\,.}{polloopperiod}
In the presence of matter in the defining representation\footnote{For
the more general twisted case, see \cite{mpw,arroyo}}
the transition functions 
$U_\mu(x)$ satisfy the  cocycle conditions \cite{tH}
\eqnl{
U_{\mu}(x)U_{\nu}(x+b_\mu)=U_\nu(x)U_\mu(x+b_\nu) \, .}{cocycle}
Under a gauge transformation the pair $(A,U)$ is mapped to 
\eqnl{ 
A^V(x)=V^{-1}(x)A(x)V(x)+i
V^{-1}(x)\d V(x)\,,\quad
U_\mu^V(x)=V^{-1}(x)U_\mu(x)V(x\!+b_\mu)\,.}{newU}
The integer-valued instanton number
\eqnl{
\qinst={1 \ov 16\pi^2}
\intl_{\T^4}\,\tr\,F\wedge 
F,}{topological}
is fully determined by the transition functions \cite{vanbaal}.
In particular, if we take all $U_\mu$ to be the
identity or equivalently a periodic gauge potential
then $\qinst=0$. Accordingly, if we are to 
describe the non-perturbative  sectors, one must consider 
non-trivial transition functions.
For a given $\qinst$ 
we only require \textit{one} set of transition functions.
If we have two sets with the same
instanton  number then they are gauge equivalent \cite{vanbaal}. 
In every instanton sector we can choose transition functions with
the following properties \cite{us}
\eqnl{
U_0=\id,\ \ U_i(x^0\!=\!0,\vx )=\id \, , \qquad i=1,2,3,\mtxt{so that}
U_i(x+b_0)=U_i(x)\,.}{condition}
The condition that $U_0=\id$ is simply the statement that the
gauge potentials are periodic in time. With \refs{polloopperiod} and 
\refs{cocycle} one obtains periodicity of $\CP(\vx)$ 
in the three spatial directions. 
The properties of the transition functions \refs{condition} 
imply that the instanton number is the winding number
of the map $\CP(\vx):\T^3\rightarrow SU(N)$, i.e.
\eqnl{
\qinst=\frac{1}{24\pi^2}\intl_{\T^3}\tr\,
(\CP^{-1}\d\CP)^3,}{polloopindex}
where $\CP=\CP(\vx )$, and $\T^3=\{x\in \T^4|x^0=0\}$. This can be 
deduced by performing the gauge transformation 
$ V(x^0,\vx )=\CP (x^0,\vx ) $ which transforms the transition functions to
$ U_0^V=\CP(\vx )\, ,\, U_i^V=\id$ and applying the well known formula
for the instanton number in terms of the transition functions \cite{vanbaal}.

Now we follow \cite{Weiss,Langmann,Lenz2,us} and seek a
gauge transformation for which the gauge transformed 
$A_0$ is time-independent and diagonal.
Consider the time-periodic gauge transformation
\eqnn{
V(x^0,\vx )=\CP(x^0,\vx )\, {\CP}^{- x^0/\beta}( \vx)\,W(\vx ),}
where $\CP(x^0,\vx )$ is the path ordered exponential \refs{defpol},
and $W(\vx)$ diagonalizes $\CP(\vx)$,
\eqnl{
\CP(\vx)=W(\vx) D(\vx) W^{-1}(\vx ).}{diagonalization}
It follows at once that the gauge transformed $A_0$,
\eqnl{
A_0^V=-{i\ov\beta}\log D(\vx),}{gfanull}
is indeed time-independent and diagonal.
Whereas $\CP(\vx)$ is smooth the factors $W(\vx)$ and
$D(\vx)$ in the decomposition \refs{diagonalization} are in
general not. The classification and implications of these
singularities are investigated below.

\begin{figure}[ht]
\begin{minipage}[ht]{16cm}
\centerline{\epsfysize=5.5 cm\epsffile{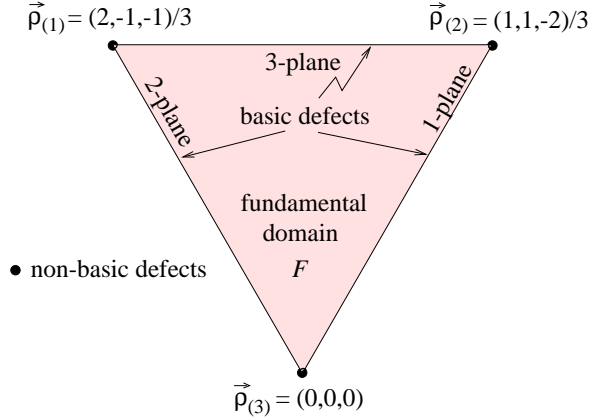}}
\caption{\label{fundomainsu3}\textsl{
The fundamental domain $\CF$ for $SU(3)$. The extremal points
of the simplex $\CF$ correspond to the center elements.
}}
\end{minipage} 
\end{figure}

The decomposition \refs{diagonalization} is not unique. 
In a first step we assign to $\CP$ a unique diagonal
\eqnl{
D(\vrho)=\hbox{diag}\Big(e^{2\pi i\rho_1},e^{2\pi i\rho_2},\dots
,e^{2\pi \rho_{N-1}},e^{2\pi i\rho_N}\Big)}{D}
in its conjugacy class. This
 is unique if we demand
\eqnl{
\vrho\in\CF,\mtxt{where}\CF=\{\vrho\in \R^N\;\vert
\rho_1 \geq \rho_2 \geq \cdots \geq \rho_N \geq \rho_1 -1,\;
\sum_{i=1}^N \rho_i = 0\}.}{ordering}
This means that the entries of $D(\vrho)$ are ordered on the circle.
The conjugacy classes are in one-to-one correspondence
with the points $\vrho$ in the \textit{fundamental domain} $\CF$.
This domain is a simplex. At its extremal points all 
but one of the $N$ inequalities in \refs{ordering} become equalities.
The extremal point where 
the only inequality is $\rho_\sigma>\rho_{\sigma+1}$, where
we have set $\rho_{N+1}\equiv \rho_1-1$, is at
\eqnn{
\vrho_{(\sigma)}=\Big(\overbrace{1,\dots,1}^\sigma,0\dots,0\Big)
-{\sigma\ov N}\Big(1,\dots,1\Big),\quad \sigma=1,\dots,N.}
The corresponding $D$ is a center element of $SU(N)$:
\eqnl{
D(\vrho_{(\sigma)})=\exp\big(2\pi i\mus\big)=
e^{-2\pi i\sigma/N}\;\id,\qquad \sigma=1,\dots,N.}{center}
We have introduced
\eqnl{
\mus=\hbox{diag}\Big(\overbrace{1,\dots,1}^{\sigma},0\dots,0\Big)
-{\sigma\ov N}\id.}{weights}
The $\{\mu_{(1)},\dots,\mu_{(N-1)}\}$ are the fundamental weights
and $\mu_{(N)}=0$.
For example, for $SU(3)$ the fundamental domain $\CF $ is 
an equilateral triangle, see fig.\ref{fundomainsu3}, 
and for $SU(4)$ a tetrahedron.
On the boundary face opposite to extremal point 
$\vrho_{(\sigma)}$ the inequality $\rho_{\sigma}\leq
\rho_{\sigma+1}$ in \refs{ordering} becomes an equality.
We call this $(N\!-\!2)$-dimensional face the $\sigma $-plane.
If $\vrho$ lies on several $\sigma$-planes,
then $D(\vrho)$ and hence $\CP$ has several coinciding
eigenvalues, see below. From now on we shall assume that the argument
$\vrho(\vx)$ of $D$ lies in the fundamental domain. Then 
\refs{diagonalization} assigns 
a unique $\vrho(\vx)$ to each $\CP(\vx)$.
We shall see that the singularities (so-called defects) in the decomposition 
\refs{diagonalization} occur at points $\vx$ for which $\vrho(\vx)$ is 
on the boundary of $\CF$.\pan
The diagonalizing matrix $W(\vx)$ in 
\refs{diagonalization} is determined only up to 
right-multiplication with an
arbitrary matrix commuting with  $D(\vx)$
\eqnl{
W(\vx)\longrightarrow W(\vx)V(\vx)\,,\quad 
V(\vx)D(\vx)V^{-1}(\vx)=D(\vx).}{residual}
At each point the residual gauge transformations $V(\vx)$
form a subgroup of $SU(N)$ which contains
the maximal Abelian subgroup $U^{N-1}(1)$ of $SU(N)$. 
At points where it is just this subgroup we can
smoothly diagonalize the Polyakov loop operator.
However, at points where it is 
\eqnl{
H \times U^l (1) \, , \quad H \, \mbox{non-Abelian} \, , \quad 
\rank(H) + l = N-1 \, ,}{centralizer}
the Polyakov loop
$\CP(\vx)$ has degenerate eigenvalues and
there are obstructions to diagonalizing
it smoothly \cite{us,tHooft,schierholz}. It
is convenient to define the \textit{defect manifold}
\eqnl{
\CD=\{\vx\in \T^3\vert \hbox{residual gauge group at }\vx
\neq U^{N-1}(1)\}}{defectmani}
on which the residual gauge symmetry is non-Abelian.
A \textit{defect} $\CD_p$ is understood to be
a connected subset of $\CD$. In the neighbourhood of a
defect the diagonalization is in general not smoothly possible
and the gauge fixing will be singular.

Now we classify the various defects arising in our gauge 
fixing. There is a defect whenever
$\CP(\vx) $ has degenerate eigenvalues, i.e. when $\vrho(\vx)$ 
is on the boundary of $\CF$. When $\vrho(\vx)$ lies on 
only one  $\sigma$-plane forming the boundary of $\CF$
then exactly two eigenvalues coincide.
We call this a  {\textit{basic}} $\sigma$-defect. 
Its residual gauge symmetry group $SU(2) \times U^{N-2}(1)$
is minimally non-Abelian. There are $N$ types of 
basic $\sigma$-defects. 
If a defect lies on several, say $k$, $\sigma$-planes,
the non-Abelian part $H$ of the 
residual gauge group has rank $ k $. $H$ is generated 
by the $SU(2)$-subgroups associated to the $\sigma$-planes 
on which the defect lies.  

Away from the defects
$W(\vx)$ in \refs{diagonalization} is unique up to a residual 
Abelian gauge transformation \refs{residual}:
\eqnl{
W(\vx)\longrightarrow W(\vx)V(\vx)\mtxt{with}
V(\vx)=e^{-i\lam(\vx)}\in U^{N-1}(1)\mtxt{on}\CD^c:=\T^3\setminus\CD
\,.}{diagonalize}
If we append to each point in $\CD^c$ the set of 
all diagonalizing matrices $W(\vx)$ we obtain a $U^{N-1}(1)$ principal bundle
over $\CD^c$. If we can find a smooth global section in this bundle 
then the diagonalization is smoothly possible outside of 
the defects, see also \cite{griesshammer}. 
To investigate the structure of the bundle we employ the
Abelian $U^{N-1}(1)$ gauge potential, $A_{mag}(\vx)$,
obtained by projecting the pure gauge $A_W=iW^{-1}dW$
onto the Cartan subalgebra,
\eqnn{
A_{mag}(\vx):=(A_W)_c\,,}
where the subscript $c$ denotes the diagonal part of $A_W(\vx)$. 
This Abelian potential is singular at the defects and on Dirac strings
joining the defects.
Under a residual gauge transformation \refs{diagonalize}
the gauge potential transforms as
\eqnn{
A_{mag}\longrightarrow A_{mag}+i(V^{-1}dV)_c=A_{mag}+d\lam \mtxt{on}\CD^c\,.}
Since $A_W$ is pure gauge, the  field strength corresponding to
$A_{mag}$
is given by
\bg\label{mfield}
F_{mag}=\d A_{mag}
=i(A_W\wedge A_W)_c\,,\eg
and it is invariant under residual Abelian gauge transformations. 

A defect may carry $N-1$ quantized
magnetic charges \cite{thooftpolyakov}. For each defect these charges form
a matrix  $Q_M$ in the Cartan subalgebra,
\eqnl{
Q_M= \frac{1}{2 \pi} \intl_{\CS} F_{mag}\, . }{magncharge}
Here $\CS$ is a surface surrounding the defect. 
The charge matrix $Q_M$ must satisfy the following quantization
condition
\eqnl{
e^{2\pi iQ_M}= \id \quad\mtxt{for each defect.}}{quantbed}
This is just the standard magnetic charge quantization
condition of  Goddard et.al \cite{go}.

If a defect $\CD_p$ divides $\CD^c$ into disconnected parts, e.g.
 a closed wall which may extend over the whole torus $\T^3$,
some comments are in order. In this case the 
surface $\CS_p$ surrounding the wall $\CD_p$
consists of several connected parts. 
If the wall defect does not extent 
over $\T^3$ every part of $\CS_p$ has no boundary and 
we get the above quantization condition, see also
\cite{Reinhardt-1098}. If the wall-defect extends
over $\T^3$ then each part of $\CS_p$ also extends over $\T^3$.
But since $F_{mag}$ is periodic on $\T^3$ 
the integral of $F_{mag}$ over each part of $\CS_p$ is again 
quantized, since we get no contributions to the magnetic charge 
from the `boundary' of the torus.  

Depending on the residual gauge symmetry in the defects
we get different types of magnetic monopoles.
This  is most elegantly expressed if we introduce the simple roots
and the lowest root,
\eqnn{
\alj=\hbox{diag}\big(\overbrace{0,\dots,0}^{j-1},1,-1,0\dots 0\big)
\;\mtxt{and}\;\al_{(N)}=\hbox{diag}\big(-1,0,\dots,0,1\big).}
The simple roots  are dual to the fundamental weights  introduced
earlier,
\eqnn{
\tr\big(\ali\cdot\muj\big)=\delta_{ij},\quad
i,j\in\{1,\dots,N\!-\!1\}.}
We have seen that to each basic $\sigma$-defect
there is an associated residual symmetry group $SU(2)$.
The root $\als$ generates the diagonal subgroup of this
$SU(2)$.
Below we shall see that a basic $\sigma$-defect
has magnetic charge
\eqnl{
Q_M=m\,\als\mtxt{with}
m\in\Z,\qquad\sigma\in\{1,\dots,N\}.}{magneticcharge}
If a defect lies on two or more $\sigma$-planes 
then $Q_M$ is an integer combination 
of the corresponding $\als$,
\eqnl{
Q_M=\sum_{\sigma=1}^{N} m_\sigma \als,\qquad m_\sigma\in\Z,
\qquad
m_\sigma=0 \, \, \mtxt{if defect is not on $\sigma$-plane.}
}{charge-on-0-plane}
The $m_\sigma$ are not overdetermined  because a defect can maximally lie 
on $N-1$ of the $\sigma$-planes. The charge matrix $Q_M$ lies
in the Cartan-subalgebra of the non-Abelian part $H$ of the 
residual symmetry group associated to the defect.

We shall also prove the following relation between the instanton 
number and the magnetic charges of defects of one type:
\eqnl{
\qinst=-\sum_{\tiny\hbox{$\sigma$-defects}} m_\sigma\, .}{instantonno}
Since \refs{instantonno} is valid for all types of defect it is immediately
obvious that for $|\qinst|>0$  \textit{every} type of defect must be present.
For example, in the case $\qinst=-1$ the total magnetic charge of a given type 
of defect is unity. The simplest (i.e. minimal)
way of achieving this would be to have exactly one monopole of each type.
One is tempted to speculate whether this minimal monopole content is
always achieved for minimal action configurations, i.e. self-dual
solutions.
Indeed, in  the recent construction of the general $\qinst=1$ caloron
solutions (i.e. instantons on $S^1 \times \R^3$) \cite{kraan}
each instanton has exactly $N$ monopole `constituents'. 
Since the magnetic field strength $F_{mag}$ lives on the compact manifold 
$\T^3$ it follows that we have overall magnetic charge neutrality:
\eqnl{
\sum_{\tiny\hbox{all defects}} Q_M = 0\,.}{charge-neutrality}
This also follows immediately from \refs{magneticcharge} and 
\refs{instantonno}; in that it
is apparent that the total magnetic charge must be proportional to
$\sum_{\sigma=1}^N \alpha_{(\sigma)}=0$.

To derive the results (\ref{magneticcharge},\ref{instantonno}) we
assume that inside a defect the residual gauge group is uniform.
This assumption is made to avoid the complication 
of `defects within defects'.
Our arguments are based on the observation that
\eqnl{
\tr\, \big(P^{-1}dP\big)^3 = 
\d\CA^{(\sigma)},\qquad\sigma=1,\ldots,N}{aab} 
where the $2$-forms are
\eqnl{
\CA^{(\sigma)} =-6\,\tr \left[A_W\wedge A_W\left(\log D -2\pi i \mu_{(\sigma)}
\right)\right] + 3\,\tr\left[A_W D^{-1}\wedge A_W D \right]\,.}{a}
These forms are smooth outside the defects, because
they are invariant under the residual Abelian gauge transformations
\refs{diagonalize}.
However, $\CA^{(\sigma)}$ can smoothly be continued into all but
the $\sigma$-defects \cite{us3}.

\begin{figure}[ht]
\begin{minipage}[ht]{16cm}
\centerline{\epsfysize=6 cm\epsffile{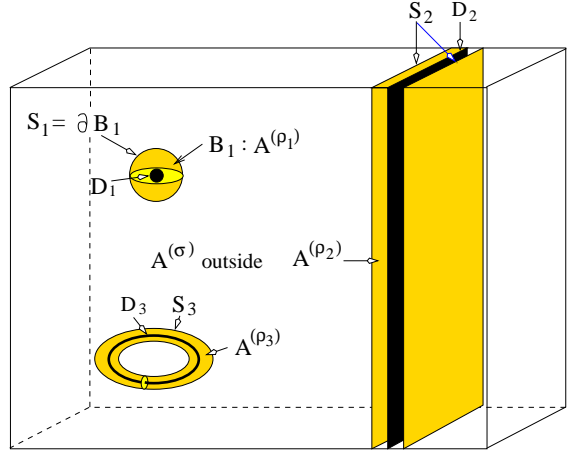}}
\caption{\label{instmon}\textsl{
We must choose two forms $\CA^{(\rho_p)}$ which are 
regular inside closed $2$-surfaces $S_p$ containing $\sigma$-defects
$D_p$.
Shown are $3$ topologically distinct defects: a point-, ring- and
wall-defect.}}
\end{minipage} \end{figure}

Now we make use of \refs{aab} to relate the magnetic charges
of the $\sigma$-defects to the instanton number. 
Away from such defects $\CA^{(\sigma)}$ is regular.
Now we surround each $\sigma$-defect with
a closed surface $\CS$ and pick a two form 
$\CA^{(\rho)}$ which is smooth
inside $\CS$, see fig.\ref{instmon}. Since a
defect can not lie on \textit{all} faces constituting
the boundary of $\CF$ there is always at least one such regular
two form with $\rho\neq \sigma$.
With (\ref{polloopindex},\ref{aab}) the instanton number reads
\eqnl{
\qinst = \frac{1}{24 \pi^2} 
\intl_{\scriptscriptstyle \rm{outside}}d\CA^{(\sigma)} +
\frac{1}{24 \pi^2} \sum_p\intl_{{\cal B}_p}d \CA^{(\rho_p)}
=\frac{1}{24 \pi^2}\sum_p\intl_{{\cal S}_p}
\big(\CA^{(\rho_p)}-\CA^{(\sigma)}\big),}{instantonnu}
where, since the $2$-forms are periodic on $\T^3$,
we get no contributions from the `boundary of the 
torus'. Using (\ref{a}) we obtain
\eqnn{
\CA^{(\rho)}-\CA^{(\sigma)}=
{12\pi i}\, \tr \Big(A_W\wedge A_W \,\big(\mu_{(\rho)}-\mus\big)\Big)\,.}
Since the magnetic field $F_{mag}$ 
is the projection to the Cartan of $iA_W\wedge A_W$ we find
\eqnl{
\CA^{(\rho)}-\CA^{(\sigma)}={12\pi} \tr\, \Big(F_{mag}\, 
\big(\mu_{(\rho)}-\mus\big)\Big)}{diff}
and end up with
\eqnl{
\qinst= \sum_{\CD_p}
\tr \Big(Q_M \,\big(\mu_{(\rho_p)}-\mus\big)\Big),}{qmag}
where we used \refs{magncharge}. The sum extends over
all $\sigma$-defects.

Let us have a closer look at the contribution 
\eqnl{
\tr \Big(Q_M \,\big(\mu_{(\rho)}-\mus\big) \Big)}{onedefect}
of a given basic $\sigma$-defect with minimal 
non-Abelian centralizer. Then all $\CA^{(\rho)}$ with $\rho\neq \sigma$ 
are regular at the defect and \refs{onedefect} must not depend on
$\mu_{(\rho)}$.
By noting that the fundamental weights are dual to the
simple roots we see at once that $Q_M$ must be proportional
to $\als$. With the quantization conditions \refs{quantbed}
we arrive at
\eqnn{
Q_M=m_\sigma\, \als \, \qquad m_\sigma\in \Z.}
With \refs{qmag} a basic $\sigma$-defect contributes $-m_\sigma$ to 
the instanton number.

A non-basic defect on the $\sigma$-plane must also lie
on one of the other boundary-planes, say the $\rho$-plane. 
Then we must not take the corresponding singular $\CA^{(\rho)}$ 
in \refs{instantonnu} or $\mu_{(\rho)}$ in \refs{onedefect}. 
We see that $Q_M$ may be an integer linear combination of 
$\als$ and $\alpha_{(\rho)}$. If the defect 
lies on several $\sigma$-planes \refs{charge-on-0-plane}
holds. The representation \refs{charge-on-0-plane} for the magnetic 
charge is unique. Note that the results 
(\ref{charge-on-0-plane},\ref{instantonno}) 
are also correct in the presence of wall defects. 

In \cite{Langmann,us} it has been shown that the (singular) gauge fixing
\eqnn{
A_0=\big(A_0(\vx)\big)_c}
can be supplemented by additional gauge fixing conditions
on the diagonal parts of $A_1,A_2$ and $A_3$. This gauge
can be achieved and is unique. One can show \cite{toappear}
that field dependent part of the Fadeev-Popov determinant associated to 
these conditions is
\eqnl{
\det\,D_0\vert_{\CH^\perp},\qquad
D_0=\pa_0-i[A_0,\,.\,].}{fpd}
Here $\CH^\perp$ is the space orthogonal to the
Cartan subalgebra.
With (\ref{gfanull},\ref{D}) our gauge fixed $A_0$ 
is diagonal and time-independent,
\eqnn{
A_0=-{i\ov\beta}\log D(\vx)=
{2\pi\ov\beta}\,\hbox{diag}\,\big(\rho_1(\vx),\dots,\rho_N(\vx)\big),}
and $\vrho(\vx)$ lies in the fundamental domain $\CF$.
With
\eqnn{
D_0\psi=D^{\xin}\pa_0\big(D^{-\xin}\psi D^{\xin}\big)D^{-\xin},\qquad
\xin=x^0/\beta.}
the eigenvalue problem $ D_0 \psi = \lambda \psi $ 
on the space of time-periodic functions results in the following simple 
equation for $ \chi = D^{-\xin}\psi D^{\xin} $:
\eqnn{
\pa_0\chi=\lam\chi,\mtxt{with boundary 
conditions}\chi(x^0+\beta)=D^{-1}(\vx)\chi(x^0)D(\vx).}
Now it is not difficult to prove \cite{Weiss,reinh2} that
\eqnn{
\det\,D_0\vert_{\CH^\perp}=C\;\prod_{1\leq i<j\leq N}
\sin^2\big\{ \pi (\rho_i(\vx)-\rho_j(\vx))\big\}}
with a field-independent constant $C$. It is just the reduced Haar measure of
$SU(N)$ \cite{Weyl}.
On the basic defects where either two $\rho$'s are equal
or $\rho_1\!-\!\rho_N=1$ the Fadeev-Popov determinant has a root of
multiplicity $2$. More generally, on a (non-basic) defect
where the residual symmetry group is $ H \times U^l(1) $ 
the determinant has a root of multiplicity $\dim(H)-\rank(H)$. 
In particular, at the center elements the determinant
has a root of multiplicity $N(N-1)$. In other words,
the multiplicity of the Fadeev-Popov determinant
at a defect is equal to the number of non-diagonal generators
of the residual gauge group at this defect. For a given $\vx$ the 
corresponding zero modes $\chi$ of $D_0\vert_{\CH^\perp}$ are just 
the elements in $H\cap\CH^\perp$.

Since the Fadeev-Popov determinant vanishes on the
boundary of $\CF$ we conclude, that the 
\textit{defects lie on the Gribov horizon}.
However, although the boundary of $\CF$ has common
points with the Gribov horizon they are not the same.
Because of the gauge fixing conditions on the
spatial components of the gauge potential the fundamental domain is
smaller (has lower dimension) than the domain bounded by the horizon.

\paragraph{Acknowledgements:}
We are grateful to Falk Bruckmann, Thomas Heinzl and Jan Pawlowski 
for helpful discussions. T.T. is indebted to Antonio Gonzalez-Arroyo 
for sharing his insights during a visit.

\end{document}